\documentclass{llncs}
\usepackage{algorithm}
\usepackage{algpseudocode} 
\usepackage{varwidth} 

\usepackage{caption}

\usepackage{amsmath,amssymb, mathtools}

\usepackage{booktabs}
\usepackage{graphicx}
\usepackage{subcaption}
\usepackage{hyperref}

\usepackage[table]{xcolor} 

\newlength{\fslength}
\makeatletter
\newcommand{\funnyP}{%
    \setlength{\fslength}{\f@size pt}%
    \reflectbox{${\mbox{P}}$}\hspace*{-.300\fslength}\mbox{${\mbox{P}}$}%
}
\newcommand{\funnyPbar}{%
    \setlength{\fslength}{\f@size pt}%
    \reflectbox{$\bar{\mbox{P}}$}\hspace*{-.300\fslength}\mbox{$\bar{\mbox{P}}$}%
}
\newcommand{\funnyPbarM}{%
    \setlength{\fslength}{\f@size pt}%
    \reflectbox{$\bar{\mbox{P}}$}\hspace*{-.300\fslength}\mbox{$\bar{\mbox{P}}_{M}$}%
}

\usepackage{stackengine}
\usepackage{array}

\usepackage{cite}

\usepackage{tabularx}

\usepackage{listings}
\usepackage{chngcntr}
\AtBeginDocument{\counterwithin{lstlisting}{section}}

\usepackage{makeidx}  
\usepackage{epsfig}
 
\usepackage{graphicx}
\usepackage[justification=centering]{caption}

\usepackage{pdfpages}

\usepackage{hyperref}
\usepackage{breakurl} 

\begin{document}

\frontmatter          
\pagestyle{headings}  
\addtocmark{Hamiltonian Mechanics} 

\mainmatter              
\title{Handling Missing Data in Performance Portability Studies}
\author{}
\institute{}
\author{Ami Marowka \orcidID{0000-0003-0914-2024}}
\institute{Parallel Research Lab, Israel\\
\email{amimar2@yahoo.com}}

\maketitle              

\begin{abstract}
\hyphenpenalty=10000
\exhyphenpenalty=10000
\tolerance=3000
\emergencystretch=3em
Missing data is a common challenge in many real-world datasets, often leading to biased results or reduced accuracy. Missing data is a {pervasive} problem in performance efficiency datasets, particularly in high-performance computing (HPC) performance portability studies. {Performance} portability scores rely on complete sets of performance {efficiencies} as inputs to their respective metrics; however, missing values can arise due to incomplete benchmarking, hardware constraints, or {implementation} gaps. Many established performance portability metrics lack mechanisms for handling missing inputs, which can result in incomplete analyses or the inability to compute scores altogether.

\smallskip
In this study, we evaluate a range of imputation methods and algorithms for addressing missing performance efficiencies. We present illustrative examples for each approach and apply them to real-world HPC {performance} portability datasets. Our results highlight the trade-offs between accuracy, robustness, and computational cost, offering practical {recommendations} for researchers seeking to mitigate the impact of missing data in performance portability assessments.

\end{abstract}

\section{Introduction}

Missing data is a common challenge across scientific disciplines, and the field of high-performance computing (HPC) performance portability is no exception. In general, missing data occurs when one or more values for a variable are absent from the dataset, either because they were not recorded or could not be obtained~\cite{Little}. This problem can substantially affect the accuracy and reliability of scientific analyses, leading to biased estimates, reduced statistical power, and, in some cases, invalid conclusions. Consequently, it is essential to identify the types and mechanisms of missingness and to apply appropriate strategies for handling them.

In the context of HPC, missing data in performance efficiency datasets poses a particular challenge for the evaluation of performance portability. Many widely used performance portability metrics cannot accommodate incomplete datasets; even a single missing value can make score computation impossible~\cite{2021-Marowka, 2023-Marowka, Deakin, Deakin2}.

For example, Figure~1 presents the architectural efficiency dataset of BabelStream Triad for an array size of $2^{29}$ FP64 elements, collected from six programming models executed on fifteen architectures~\cite{Little}. Here, approximately 39\% of the entries are missing, and each programming model has at least one missing value. As a result, performance portability scores computed with the {\funnyP} metric are zero for all cases---because it assigns a score of zero whenever any platform is unsupported. By contrast, the {\funnyPbar} metric, which is based on the arithmetic mean and considers only supported platforms, produces meaningful scores~\cite{2023-Marowka}. However, a nonzero {\funnyPbar} score may still be based on a substantially incomplete subset of platforms; for example, the OpenCL score in this dataset is computed from only 6 of the 15 efficiencies. Imputing estimable missing entries can therefore provide a more representative basis for the aggregate score and reveal information that is otherwise absent from the comparison.

Several factors can lead to missing entries in HPC performance datasets~\cite{Deakin, Deakin2}:

\begin{enumerate}
    \item Infeasible platform--programming model combinations (e.g., CUDA on non-NVIDIA hardware).
    \item Runtime or measurement errors during result collection.
    \item Lack of compiler or runtime support for certain programming models.
    \item Insufficient memory for a specific problem size on a given platform.
    \item Build-time errors that cannot be resolved.
    \item Runtime crashes or erroneous outputs.
\end{enumerate}

Data imputation---the process of replacing missing values with estimated substitutes---is a widely used approach to address this issue~\cite{Buuren, Murray}. Imputation methods range from simple statistical techniques to advanced machine learning algorithms. 

In HPC performance efficiency datasets, there is no universally optimal imputation method. The suitability of a given approach depends on the specific patterns of missingness, the underlying data structure, and the intended application. Empirical evaluation is therefore essential to identify the best approach for a given case.

In this work, we systematically evaluate three imputation techniques applied to real-world HPC performance portability datasets. We illustrate each method, compare their accuracy and robustness, and assess computational costs. Our contributions include:

\begin{itemize}
    \item \textbf{Statistical methods} --- e.g., mean imputation.
    \item \textbf{Predictive modeling} --- e.g., linear regression imputation.
    \item \textbf{Matrix factorization methods} --- including low-rank matrix factorization (LRMF) via singular value decomposition (SVD).
\end{itemize}

The results provide practical guidance for selecting imputation strategies that preserve the integrity of performance portability analyses while balancing accuracy, robustness, and computational feasibility.

We focus on reconstruction-oriented imputation methods suitable for small, structured efficiency tables. Classical likelihood-based and fully Bayesian missing-data frameworks are designed for statistical inference and hypothesis testing and are therefore outside the scope of this work.


\begin{figure}[!htb]

    \captionsetup{justification=centering} 
     \centering
     \includegraphics [height=7cm, width=0.9\linewidth]{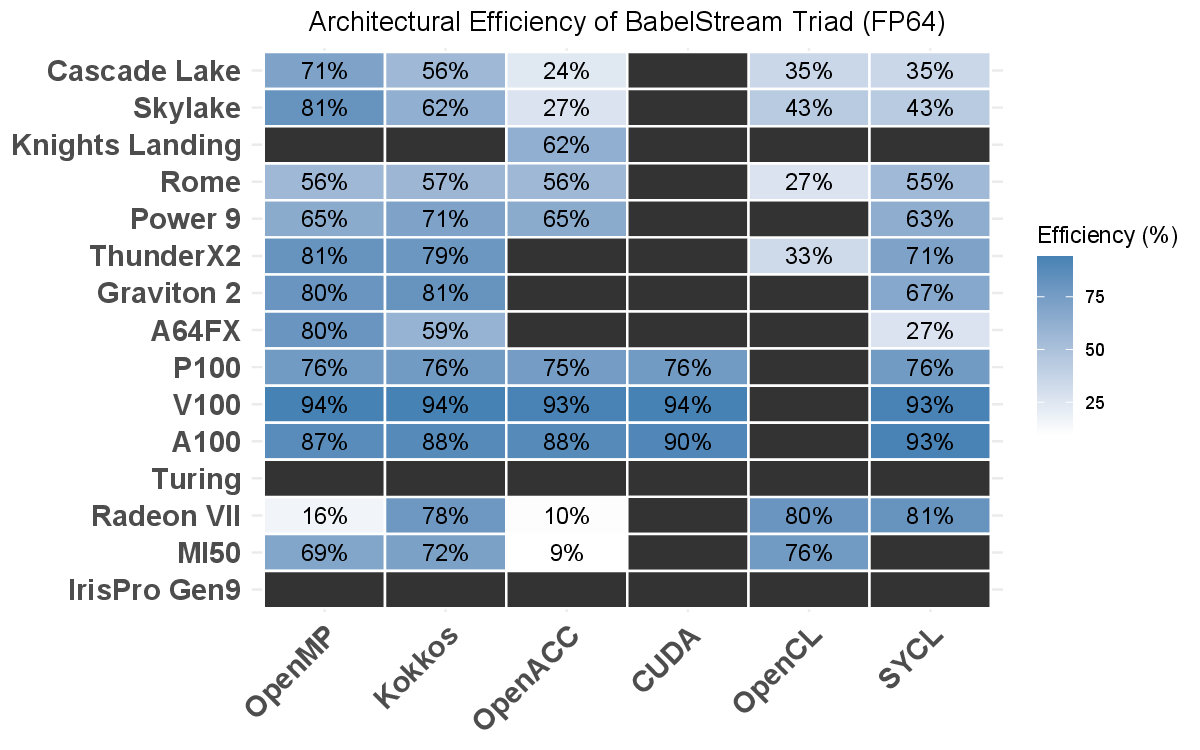}
\label{fig:fig1}  

\caption{BabelStream Triad Architectural Efficiencies for arrays of length $2^{29}$ FP64 elements 
              [From \cite{Deakin}]. 39\% of the data is missing.}


\vspace{0.6em}

\centering
\captionsetup{type=table}
\resizebox{0.8\textwidth}{!}{%
\begin{tabular}[htbp]{lcccccc} 
\toprule
  \multicolumn{6}{c}{\textbf{Performance Portability}} \\
\midrule
 & \textbf{OpenMP} & \textbf{ Kokkos} & \textbf{ OpenACC} & \textbf{ CUDA} & \textbf{ OpenCL} & \textbf{ SYCL} \\
\midrule
\rowcolor{red!65}
{\funnyP}   & {\bf 0\%} & {\bf 0\%} & {\bf 0\%} & {\bf 0\%} & {\bf 0\%} & {\bf 0\%} \\
{\funnyPbar} & 71.33\% & 72.25\% & 50.90\% & 86.66\% & 54.85\% & 63.54\% \\
\bottomrule
\end{tabular}
}
\caption{BabelStream Triad Performance Portability Scores.} 

\vspace{0.6em}
\hfill

\centering
\captionsetup{type=table}
\begin{tabular}{lccc}
\multicolumn{4}{c}{} \\ 
\midrule
\textbf{} & \textbf{(P100)} & \textbf{(V100)} & \textbf{(A100)}  \\
\midrule
OpenMP    & 76\%  & 94\%  & 87\%   \\
Kokkos & 76\%  & 94\%  & 88\%    \\
OpenACC  & 75\%  & 93\%  & 88\%   \\
\bottomrule
\end{tabular}
\caption{A sample of BableStream Triad Architectural Efficiencies  \\ (From Figure 1.)}
\end{figure}


\section{Benchmarking Methodology}

In the following sections, we describe the imputation methods under study using a real architectural efficiency dataset drawn from a evaluation of GPU architectures deployed in modern supercomputers (Table~2). 

The controlled \(3\times3\) examples are intended to demonstrate and validate the imputation procedure rather than to benchmark matrix-completion algorithms comprehensively. Performance-portability efficiency matrices are generally small, since their dimensions correspond to implementations and platforms. A systematic comparison across methods, masking ratios, and missingness patterns is left for an extended study.




For demonstration purposes, we assume that one or more performance efficiency values are missing, and compare the imputed values produced by different models with the actual measured values. This raises an important question: \textit{If we compare an imputed value to the measured value, what does that comparison truly signify, and is it meaningful?}

Benchmarking measurements are inherently subject to noise. The ``true'' value is itself variable due to factors such as system noise and benchmarking instability. If the imputed value closely matches the measured value, we consider the imputation model to have low error---assuming that the measured value serves as a reasonable proxy for ground truth. Conversely, large discrepancies may arise from poor model assumptions, high data variance, measurement noise, or outliers in the observed dataset.

Despite these limitations, such comparisons are standard practice when evaluating imputation accuracy. A common approach is cross-validation masking: selectively hide known entries, perform imputation, and compare the predicted values against the actual observed values \cite{Jager,Woznica}.

To assess imputation quality, we employ standard regression-based metrics: Root Mean Squared Error (RMSE), Mean Absolute Error (MAE) and Mean Absolute Percentage Error (MAPE). These evaluations do not establish an absolute truth; rather, they assess how well the imputation model reproduces observed behavior in the presence of uncertainty.

\begin{equation}
\scalebox{0.85}{$
\begin{aligned}
\mathrm{RMSE} &= \sqrt{\frac{1}{n}\sum_{i=1}^{n}\left(\hat{y}_i - y_i\right)^2}, \quad
\mathrm{MAE}  = \frac{1}{n}\sum_{i=1}^{n}\left|\hat{y}_i - y_i\right|, \quad
\mathrm{MAPE} = \frac{100}{n}\sum_{i=1}^{n}\left|\frac{\hat{y}_i - y_i}{y_i}\right|.
\end{aligned}
$}
\end{equation}

From an analytical perspective, each metric highlights a complementary aspect of imputation quality. RMSE emphasizes larger deviations by squaring the error, making it particularly sensitive to outliers and useful for identifying methods that occasionally produce large reconstruction errors. MAE provides a more robust, scale-consistent measure of typical error magnitude and reflects the average absolute discrepancy between imputed and true values. MAPE normalizes the error relative to the magnitude of the true entries, enabling comparison across entries with different scales and offering an interpretable, percentage-based assessment of relative accuracy. Taken together, these metrics allow us to distinguish between methods that minimize worst-case errors, those that achieve consistently small deviations, and those that preserve relative accuracy across heterogeneous efficiency values.



Finally, because the dataset consists of architectural efficiency values, it is often more appropriate to use statistical methods (e.g., mean imputation) or predictive models (e.g., linear regression) along a single dimension---namely, the programming model dimension. This is due to the fact that architectural efficiency is computed relative to an architecture-specific baseline (theoretical peak performance or roofline peak performance); thus, the denominator varies between architectures but remains constant within a given architecture.

\section{Simple Statistical Imputation --- Mean Imputation}

Mean imputation is a simple statistical technique for handling missing data by
replacing each missing value with the mean of the available observations in the
same variable (e.g., column or row). In the context of architectural-efficiency
datasets, this involves calculating the average efficiency across all observed
entries for a specific hardware platform and substituting this mean for any
missing entries in that column. The method is computationally inexpensive, easy
to implement, and well suited for small numerical datasets with limited missing
data.

However, mean imputation assumes that missing values are missing completely at
random (MCAR) and ignores relationships between variables, potentially
introducing bias. Each column is treated independently, and inter-column
correlations are not leveraged. Moreover, the presence of extreme values or
heterogeneous behavior across programming models can skew the mean, leading to
inaccurate imputations and reduced model fidelity \cite{Little}. For these
reasons, mean imputation is commonly used as a baseline rather than as a
standalone solution in performance-analysis studies.

\begin{table}[t]
\centering
\caption{Mean imputation results for BabelStream architectural efficiencies
(Table~2). Imputation is performed column-wise using
the mean of observed entries.}
\label{tab:mean_impute_babelstream}
\begin{tabular}{lccccc}
\toprule
\textbf{Mask} & \textbf{Actual (\%)} & \textbf{Imputed (\%)} &
\textbf{RMSE} & \textbf{MAE} & \textbf{MAPE (\%)} \\
\midrule
OpenMP--P100   & 76.0 & 75.5 & 0.5 & 0.5 & 0.66 \\
Kokkos--V100   & 94.0 & 93.5 & 0.5 & 0.5 & 0.53 \\
OpenACC--A100  & 88.0 & 87.5 & 0.5 & 0.5 & 0.57 \\
\bottomrule
\end{tabular}
\end{table}

\section*{Discussion}

The mean-imputation results for the three held-out entries---OpenMP--P100,
Kokkos--V100, and OpenACC--A100---exhibit uniformly small errors, with RMSE and
MAE equal to $0.5$ percentage points in all cases and relative errors (MAPE)
below $1\%$. These outcomes illustrate several important properties of mean
imputation when applied to small HPC architectural-efficiency tables:

\begin{enumerate}

  \item \textbf{Column stability drives accuracy.}

  \begin{itemize}
    \item \emph{P100}: The column values (76, 76, 75) are tightly clustered,
    yielding a mean of $75.5\%$. The imputed OpenMP value lies close to the true
    efficiency of $76\%$.

    \item \emph{V100}: Efficiencies are highly stable (94, 94, 93), producing a
    mean of $93.5\%$, which closely approximates the held-out Kokkos entry.

    \item \emph{A100}: With values (87, 88, 88), the column exhibits minimal
    dispersion, making the column mean ($87.5\%$) a reliable proxy for the
    missing OpenACC value.
  \end{itemize}

  \item \textbf{Sensitivity to dispersion remains a concern.}

  Although mean imputation performs well here, this success is largely due to
  the low variance within each column. In columns with higher dispersion or
  pronounced outliers, the mean may be a poor representative of individual
  entries, leading to significantly larger errors. Thus, low error in these
  cases should not be generalized to more heterogeneous efficiency tables.



  \item \textbf{Practical guidance for HPC efficiency datasets.}

  \begin{itemize}
    \item Mean imputation is most appropriate when column-wise efficiencies are
    stable and unimodal.

    \item Columns should be screened for dispersion (e.g., variance or
    interquartile range) before applying mean imputation.

    \item When variability is higher or missingness is systematic, more robust
    or structure-aware methods---such as low-rank matrix completion---are preferable.

    \item Imputation uncertainty should be quantified or contextualized when
    imputed values are used in aggregated performance or portability metrics.
  \end{itemize}

\end{enumerate}



\section{Linear Regression Imputation}

Linear regression imputation is a predictive approach that estimates missing
values by exploiting linear trends in the observed data. Rather than replacing
a missing entry with a constant summary statistic (such as the mean), regression
imputation models how the response variable varies as a function of an ordered
predictor and extrapolates this relationship to the missing location.

Conceptually, this corresponds to fitting a straight line through the observed
points and evaluating the fitted model at the position of the missing entry (Figure 2).
Compared with mean imputation, linear regression is more context-aware, as it
captures monotonic trends or gradients in the data.

Formally, a simple linear model
\begin{equation}
\hat{y} = \beta_0 + \beta_1 x
\end{equation}
is fit to the observed data using ordinary least squares (OLS), which selects
$(\beta_0,\beta_1)$ to minimize the sum of squared residuals
$\sum_i (y_i - \hat{y}_i)^2$. The fitted coefficients are then used to predict the
missing response $\hat{y}$ at the corresponding predictor value $x$.

In the architectural-efficiency setting of Table~2,
the predictor $x$ is the index assigned to the programming models, ordered here as
OpenMP $\rightarrow$ Kokkos $\rightarrow$ OpenACC, and the response $y$ is the
observed efficiency for a fixed GPU architecture (column). This indexing is a
simplifying assumption rather than a physically meaningful predictor: changing
the programming-model ordering changes the assigned $x$ values and can therefore
change the imputed result. Consequently, regression across model indices should
be regarded only as a lightweight illustrative baseline unless an independently
justified ordering or predictive feature is available.

\subsection{Demonstration Setup}

We evaluate linear regression imputation on three single-entry masking scenarios
derived from Table~2---OpenMP--P100,
Kokkos--V100, and OpenACC--A100

For each case, an OLS regression line is fit to the remaining two observed
efficiencies in the corresponding GPU column, indexed by the ordered
programming models. The fitted model is then evaluated at the missing index.
Imputation accuracy is assessed by comparing the predicted value to the held-out
true value using RMSE, MAE, and MAPE.

\begin{table}[t]
\centering
\caption{Linear regression imputation results for the BabelStream architectural
efficiency matrix (Table~2). For each case, a single
observed entry is held out and imputed via ordinary least squares regression
along the ordered programming models.}
\label{tab:linreg_impute_babelstream}
\begin{tabular}{lccccc}
\toprule
\textbf{Mask} &
\textbf{Actual (\%)} &
\textbf{Imputed (\%)} &
\textbf{RMSE} &
\textbf{MAE} &
\textbf{MAPE (\%)} \\
\midrule
OpenMP--P100   & 76.0 & 77.0 & 1.0 & 1.0 & 1.32 \\
Kokkos--V100   & 94.0 & 93.5 & 0.5 & 0.5 & 0.53 \\
OpenACC--A100  & 88.0 & 89.0 & 1.0 & 1.0 & 1.14 \\
\bottomrule
\end{tabular}
\end{table}

\begin{figure}[t]
\centering
\begin{subfigure}{0.48\linewidth}
  \centering
  \includegraphics[width=1.0\linewidth]{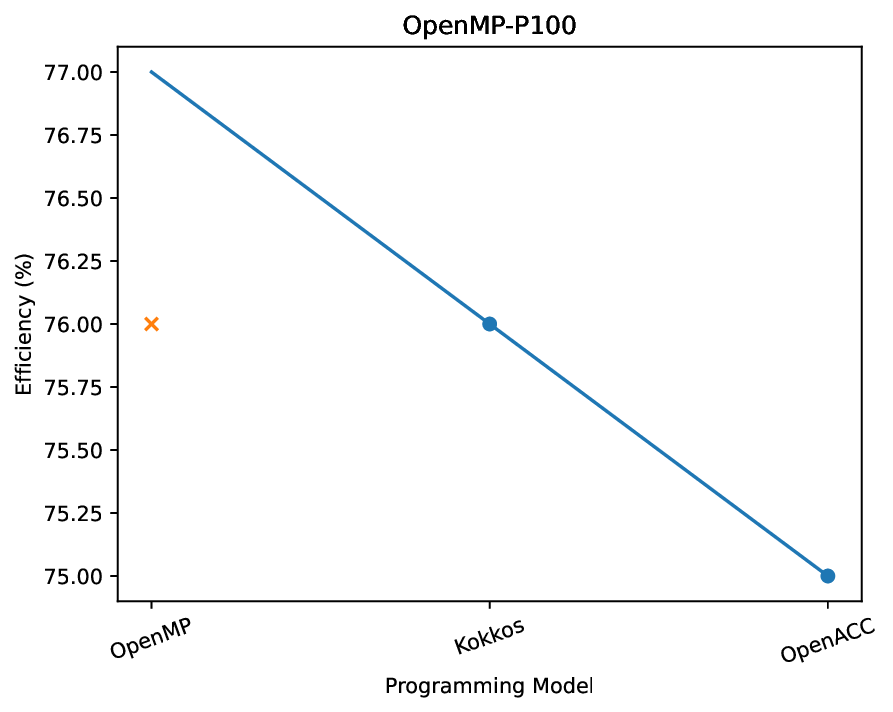}
  \caption{P100 (OpenMP held-out)}
\end{subfigure}
\begin{subfigure}{0.48\linewidth}
  \centering
  \includegraphics[width=1.0\linewidth]{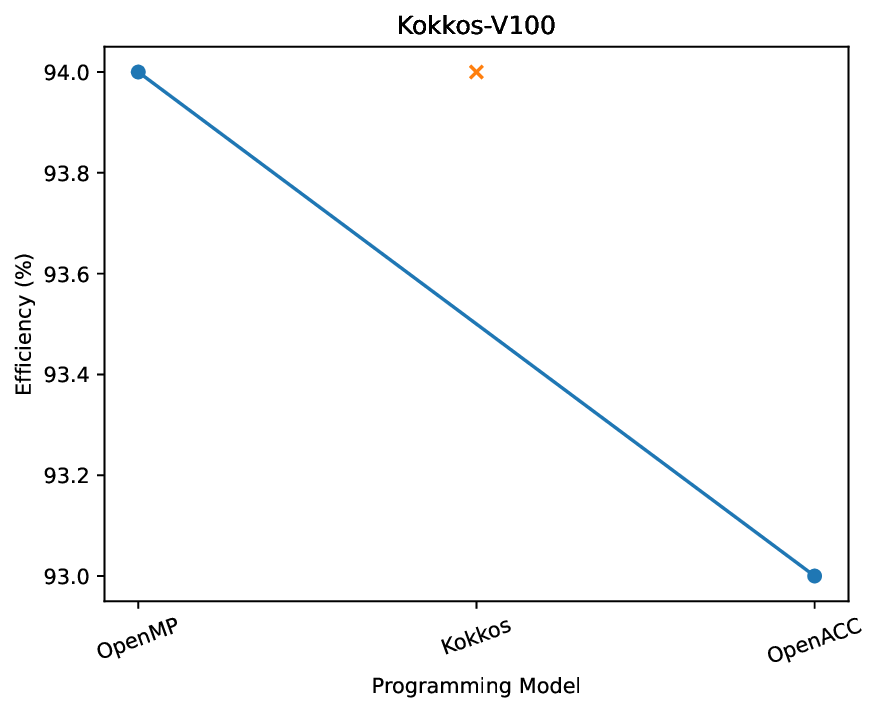}
  \caption{V100 (Kokkos held-out)}
\end{subfigure}

\caption{Linear regression imputation across programming models.}
\label{linear_regression}
\end{figure}

\subsection{Results and Interpretation}

Linear regression imputation yields small reconstruction errors across all three
cases, with absolute errors between $0.5$ and $1.0$ percentage points (Table 4). The lowest
error is observed for Kokkos--V100 (MAPE $=0.53\%$), reflecting the strong linear
trend and low dispersion in the V100 column. Larger errors occur for OpenMP--P100
and OpenACC--A100, where the true values deviate slightly from the extrapolated
linear trend.

These results highlight the strengths and limitations of regression-based
imputation. When an ordering is independently meaningful and efficiencies vary
smoothly and approximately linearly along it, regression can incorporate trend
information that mean imputation ignores. In the present demonstration, however,
the programming-model index has no intrinsic predictive meaning, so the result is
order-dependent. Moreover, with only a small number of observed points, the fitted
line can be sensitive to endpoint values, and extrapolation errors may occur when
the true behavior deviates from linearity.

Overall, linear regression imputation provides a lightweight, context-aware
baseline that bridges the gap between purely statistical methods (mean
imputation) and multivariate, structure-aware approaches such as low-rank matrix
completion.
 



\section{Matrix Completion Methods}

The matrix completion problem concerns determining whether a partially observed matrix can be completed so as to satisfy a prescribed property \cite{Laurent}. 

Formally, given a partially observed matrix \(X \in \mathbb{R}^{m \times n}\) with observed index set \(\Omega \subseteq \{1,\dots,m\}\times\{1,\dots,n\}\), the aim of \emph{matrix completion} is to infer the missing entries \(X_{ij}\) for \((i,j)\notin \Omega\) by exploiting structure in the observed data.

A common and highly effective assumption is {\bf low rank}: the data can be well-approximated by a small number of latent factors. 
In other words, the idea is to guess what those missing values should be based on the patterns found in the known ones. But it does this using the presumption that even though the matrix looks complex and incomplete, the patterns behind it are actually simple. They can be explained by a small number of hidden traits or factors.
Therefore, the algorithms that we used here assume:
\begin{itemize}
\item {Each column (e.g., a platform) is a mix of a few latent capabilities.}
\item {Each row (e.g., a programming model) is a mix of a few latent demands.}
\item {The observed numbers are just a match score between those capabilities and demands.}
\end{itemize}

This is the essence of low-rank matrix factorization---the belief that even a big matrix is built from just a few underlying building blocks.
Under this view, we aim to find a low-rank matrix that agrees with the known entries and generalizes sensibly to the missing ones. While direct low-rank factorization leads to a non-convex optimization landscape with potentially many local minima, carefully designed algorithms are remarkably effective in practice and, under standard assumptions, can recover the ground truth with high probability \cite{Goldfarb}.

A widely effective modeling choice assumes that \(X\) is well approximated by a low-rank matrix.
Concretely, we posit \(X \approx U V^{\top}\), where \(U\in\mathbb{R}^{m\times k}\) and \(V\in\mathbb{R}^{n\times k}\) with \(k \ll \min(m,n)\).
In an efficiency table (e.g., rows = programming models, columns = platforms), each column can be represented by a small number of ``platform traits'' and each row by a few ``model traits''; the entries arise from their interactions.
This explains why a small rank \(k\) can capture most of the signal even with many missing entries.

Matrix completion is NP-hard in the worst case, yet becomes tractable under informed conditions---e.g., approximate low rank, incoherence of singular vectors, and sufficiently well-distributed observations. In such regimes, efficient algorithms can provide accurate reconstructions with provable guarantees.

Matrix completion methods are widely used across domains where missing data are common and latent low-rank structure is a reasonable assumption, including {\bf machine learning \& recommender systems} (e.g., Netflix, Amazon, Spotify, YouTube), 
{\bf genomics/bioinformatics} (imputing gaps in microarray and \mbox{RNA-seq} data for downstream analysis).


\subsection{Common Iterative Low-Rank Matrix Completion Algorithm}
\label{sec:global_algorithm}

We write \(\|\cdot\|_F\) for the Frobenius norm and use \(\Omega\subseteq\{1,\dots,m\}\times\{1,\dots,n\}\) to denote the set of observed indices.
The Frobenius norm of a matrix $A \in \mathbb{R}^{m \times n}$ is defined as
\scalebox{0.8}{$
\lVert A \rVert_{F,\Omega}
= \sqrt{\sum_{(i,j)\in\Omega} a_{ij}^2}
$}. 

\begin{algorithm}[t]
\caption{Common Iterative Low-Rank Matrix Completion Algorithm}
\label{alg:global}
\begin{algorithmic}[1]
\Require
\begin{varwidth}[t]{\linewidth}
\(X \in \mathbb{R}^{m\times n}\) partially observed on \(\Omega\subseteq\{1,\dots,m\}\times\{1,\dots,n\}\), \\
\(\mathcal{P}_\Omega\) the projection that keeps entries on \(\Omega\) and zeros elsewhere, target rank $k$
\end{varwidth}

\Ensure
Completed matrix $\hat X$

\State \textbf{Initialize} missing entries to obtain a fully specified matrix $\hat X^{(0)}$
(e.g., row/column means or a simple constant); select a target rank $k$.  

\For{$t = 0,1,2,\ldots$}  
\State 
\begin{varwidth}[t]{\linewidth}
\textbf{Factorize} the current iterate with target rank \(k\) to extract the dominant structure.\\
\hspace*{1cm}Factors of SVD method:  $U_k ,\Sigma_k , V_k^{\top}$ 
\end{varwidth}
\State 
\begin{varwidth}[t]{\linewidth}
\textbf{Reconstruct} a low-rank estimate \(\tilde X^{(t)}\) from the factors. \\
\hspace*{1cm}SVD method:  \(\tilde X^{(t)} = \) $U_{k}\Sigma_{k}V_{k}^{\top}$ 
\end{varwidth}        
\State 
\begin{varwidth}[t]{\linewidth}
\textbf{Update} only the missing entries:
    \[
    \hat X^{(t+1)}
    =
    \mathcal{P}_{\Omega}(X)
    +
    \mathcal{P}_{\Omega^c}\!\left(\tilde X^{(t)}\right)
    \]

Here, $\Omega^c$  denotes the set of unobserved (missing) matrix entries.\\
Observed entries remain unchanged. 
\end{varwidth}

\State
\begin{varwidth}[t]{\linewidth}
\textbf{Check convergence} and stop when the relative change is small threshold or a maximum number of iterations is reached. In other words, we measure the difference between the current iteration and the previous iteration for the missing values using Frobenius norm.
\[
{\bigl\|\hat X_{\Omega^c}^{(t+1)}-\hat X_{\Omega^c}^{(t)}\bigr\|_F} 
\;<\; \varepsilon.\]
\end{varwidth}
\EndFor
\State \Return $\hat X$
\end{algorithmic}
\end{algorithm}

Algorithm~\ref{alg:global} presents the common iterative structure used throughout
this work for completing missing entries in performance-efficiency matrices.
The algorithm operates on a fully specified working matrix $\hat X^{(t)}$,
which is updated iteratively.

At each iteration, the algorithm consists of three conceptual steps.
First, a rank-$k$ low-rank approximation of the current matrix $\hat X^{(t)}$
is computed (Line~3). 

Second, the reconstructed low-rank approximation $\tilde X^{(t)}$ (Line~4) is used
to update only the missing entries, while the originally observed values
are kept unchanged (Line~5). This consistency enforcement ensures that
the algorithm does not alter measured data.

Finally, the procedure is repeated until a convergence criterion is met (Line~6),
typically defined in terms of stabilization of the imputed values or
changes in the reconstructed matrix.


Conceptually, Algorithm~\ref{alg:global} alternates between
(i) extracting a low-rank structure from the current estimate of the matrix
and (ii) re-imposing consistency with the observed data.
This separation allows different low-rank models to be compared
within an identical iterative completion framework.


\subsection{Initialization of Algorithm~1}
\label{subsec:initialization}

The initialization step of Algorithm~1 establishes a fully specified starting point for the iterative factorization procedure and fixes the structural parameters that remain constant throughout the algorithm.

\paragraph{Initialization of missing entries.}
Given an incomplete data matrix $X$ with observed index set $\Omega$, the algorithm first constructs an initial fully specified matrix $\hat X^{(0)}$ by assigning values to the missing entries. This initialization serves only as a starting point and does not constrain the final solution.

In practice, simple deterministic strategies are sufficient, such as replacing missing entries by row means, column means, or a global constant (e.g., the overall mean of observed values). These choices provide a reasonable baseline estimate while avoiding the introduction of strong structural assumptions. Subsequent iterations overwrite the initialized values using low-rank approximations derived from the factorization method, ensuring that the influence of the initial fill diminishes as the algorithm converges.

\paragraph{Choice of target rank $k$.}
The target rank $k$ controls the dimensionality of the latent representation and is fixed for the entire execution of Algorithm~1. Once selected, $k$ is not adapted dynamically, even if intermediate decompositions of the current iterate $\hat X^{(t)}$ suggest a lower effective rank.


\paragraph{Example: rank selection via reconstruction accuracy.}
A practical and widely applicable approach for selecting $k$ is based on reconstruction accuracy using observed data. Specifically, a small subset of observed entries is temporarily masked and treated as validation data. For a range of candidate ranks $k \in \{k_1,\ldots,k_L\}$, Algorithm~1 is executed using only the remaining observed entries, and the reconstruction error on the masked values is evaluated.

The rank $k$ that minimizes this validation error---or achieves a satisfactory trade-off between accuracy and model complexity---is then selected and fixed for the full execution of the algorithm. This approach is particularly well suited to small performance-efficiency matrices, where exhaustive testing over a modest range of ranks is computationally inexpensive and yields stable rank choices.





\section{Singular Value Decomposition (SVD)}
\label{sec:svd}

\paragraph{Definition and basic properties.}
For a matrix \(A \in \mathbb{R}^{m \times n}\), a (full) singular value decomposition (SVD) factors \(A\) as
\[
A \;=\; U \, \Sigma \, V^{\top},
\]
where \(U \in \mathbb{R}^{m\times m}\) and \(V \in \mathbb{R}^{n\times n}\) are orthogonal matrices, and \(\Sigma \in \mathbb{R}^{m\times n}\) is diagonal with nonnegative entries (the singular values) ordered from largest to smallest (Figure 3). The columns of \(U\) are the left singular vectors; the columns of \(V\) (equivalently, the rows of \(V^{\top}\)) are the right singular vectors \cite{Golub}.
\paragraph{Existence and computational cost.}
Every real matrix admits an SVD \cite{Trefethen}.
Standard dense algorithms (e.g., Golub-Reinsch variants) are heavily optimized; a typical complexity bound is on the order of the smaller of \(O(m^2 n)\) and \(O(n^2 m)\) for computing a full SVD \cite{Reinsch}.

\paragraph{Interpretation.}
Writing \(\Sigma=\mathrm{diag}(\sigma_1,\dots,\sigma_r)\) with \(\sigma_1\ge\dots\ge\sigma_r\ge 0\) and \(r=\mathrm{rank}(A)\), the SVD exposes orthonormal directions \(v_i\) in the domain (columns of \(V\)) that are mapped to orthonormal directions \(u_i\) in the domain (columns of \(U\)) with gains \(\sigma_i\):
\(
A v_i = \sigma_i u_i
\).
This makes SVD central to low-rank modeling, denoising, and matrix completion.

\begin{figure}[t]
  \centering
  \scalebox{0.8}{
  \includegraphics[width=\linewidth]{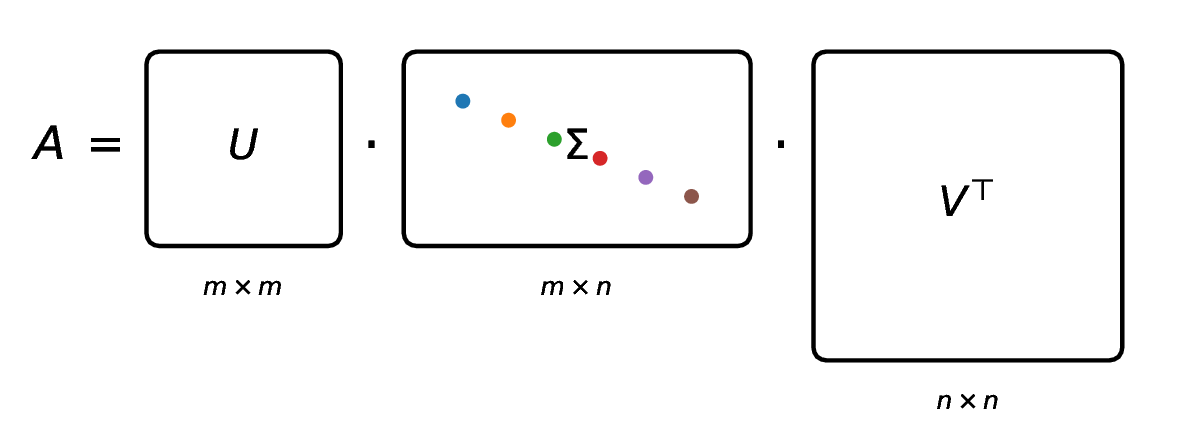}
   }
   \caption{Block diagram of the SVD \(A=U\Sigma V^{\top}\): \(U\) and \(V\) are orthogonal, and \(\Sigma\) is diagonal with nonnegative, descending singular values. The   shapes shown reflect \(U\in\mathbb{R}^{m\times m}\), \(\Sigma\in\mathbb{R}^{m\times n}\), and \(V^{\top}\in\mathbb{R}^{n\times n}\).}
  \label{fig:svd-block}
\end{figure}


\subsection{Low-Rank Approximations from the SVD}
\label{sec:svd-lowrank}
 
\paragraph{Setup.}
Let \(A \in \mathbb{R}^{m\times n}\) with singular value decomposition (SVD) 

\[
  A \;=\; U\,\Sigma\,V^{\top},
\]
where \(U\in\mathbb{R}^{m\times m}\) and \(V\in\mathbb{R}^{n\times n}\) are orthogonal, and
\(\Sigma=\mathrm{diag}(\sigma_1,\ldots,\sigma_r)\in\mathbb{R}^{m\times n}\) with singular
values \(\sigma_1 \ge \sigma_2 \ge \cdots \ge \sigma_r > 0\) and \(r=\mathrm{rank}(A)\).
Denote by \(U_k\in\mathbb{R}^{m\times k}\), \(\Sigma_k\in\mathbb{R}^{k\times k}\), and
\(V_k\in\mathbb{R}^{n\times k}\) the leading \(k\) left/right singular vectors and singular values.

\paragraph{Simplest rank-\(k\) approximation (truncated SVD).}
The standard procedure to obtain a rank-\(k\) approximation of \(A\) is:
\begin{enumerate}
  \item Compute (or access) the SVD \(A=U\Sigma V^\top\).
  \item Keep only the top \(k\) singular values and the associated singular vectors:
        \(U_k=[u_1,\dots,u_k]\), \(V_k=[v_1,\dots,v_k]\), \(\Sigma_k=\mathrm{diag}(\sigma_1,\dots,\sigma_k)\).
  \item Form the truncated reconstruction
        \[
          A_k \;=\; U_k \,\Sigma_k\, V_k^{\top}.
        \]
\end{enumerate}
Intuitively, \(A_k\) captures the dominant \(k\) ``modes'' (largest singular values) of \(A\).


\section{Benchmarking Algorithm~1}
\label{sec:benchmarks-small}

To illustrate the practical behavior of Algorithm~1, we consider a small
$3\times 3$ architectural-efficiency matrix derived from BabelStream,
shown in Table~2. Entries represent architectural
efficiency (in percent) for three programming models and three GPU architectures.

Four masking scenarios are considered:
(i) OpenMP--P100,
(ii) Kokkos--V100,
(iii) OpenACC--A100,
and (iv) all three entries masked simultaneously.
All benchmarks were implemented in Python using NumPy for linear algebra.
SVD was computed via \texttt{numpy.linalg.svd}. 

All experiments are run for ranks $k\in\{1,2\}$. Missing entries are initialized
using column means computed from observed values. The outer iteration stops
when the maximum absolute change in masked entries falls below $10^{-6}$,
or when a maximum of 1000 iterations is reached.
 
\begin{table}[t]
\centering
\caption{Matrix Completion imputation results for the BabelStream architectural
efficiency matrix (Table~2). For each case, a single
observed entry is held out and imputed via iterative low-rank matrix completion using SVD.}
\label{tab:mask-a}
\begin{tabular}{lcccccc}
\hline
Mask & Actual (\%) & Imputed (\%) & RMSE & MAE & MAPE (\%) & \# Iter. \\
\hline
OpenMP--P100   &  76.00 & 76.20 & 0.0020 & 0.0020 & 0.27 & 74 \\
Kokkos--V100   &  94.00 & 94.11 & 0.0011 & 0.0011 & 0.12 & 23 \\
OpenACC--A100  &  88.00  & 88.40 & 0.0040 & 0.0040 & 0.46 & 1000 \\
\hline
\end{tabular}
\end{table}

\begin{table}[t]
\centering
\caption{Mask all three entries simultaneously.}
\label{tab:mask-d}
\begin{tabular}{ccc|ccc}
\hline
 \multicolumn{3}{c}{Imputed Values (\%)} & \multicolumn{3}{c}{Error Metrics} \\
 OpenMP & Kokkos & OpenACC & RMSE & MAE & MAPE (\%) \\
 --P100 & --V100 & --A100 &   &   &     \\
\hline
 76.20 & 94.11 & 88.40 & 0.0081 & 0.0066 & 0.79 \\
\hline
\end{tabular}
\end{table}

Tables~\ref{tab:mask-a} report reconstruction accuracy for
the three single-entry masking scenarios using $k=2$. Reported metrics are
computed only on the masked entries.
Table~\ref{tab:mask-d} reports results when all three diagonal entries are
masked simultaneously. In addition to aggregate error metrics, we explicitly
list the individual imputed values, highlighting how different methods
redistribute error across entries. 

Several observations emerge. First, truncated SVD provides stable and
competitive reconstructions, but may still reach the maximum number of outer
iterations because convergence is governed by changes in masked entries rather
than by the optimality of the rank-$k$ approximation itself.

Finally, repeating the benchmarks with $k=1$ (not shown) yields qualitatively
similar trends but with larger reconstruction error, confirming that $k=2$
provides a better balance between bias and variance for this matrix.



\section{Conclusions}

Missing data is a pervasive challenge in performance-efficiency datasets used
for evaluating performance portability in high-performance computing.
Because many established portability metrics require complete inputs, even a
small fraction of missing values can invalidate or severely distort the resulting
analysis.

In this work, we systematically examined several imputation strategies for
handling missing architectural-efficiency values, ranging from simple
statistical methods to predictive modeling and low-rank matrix completion.
Using real-world HPC datasets, we demonstrated how each method trades off
simplicity, robustness, accuracy, and computational cost. Our results show
that while simple approaches such as mean or linear regression imputation can
be effective for small, low-variance tables, low-rank matrix completion methods
provide a more flexible and structure-aware solution when missingness is more
severe or heterogeneous.

The present workshop paper focuses on demonstrating the applicability of the common
iterative low-rank procedure rather than on an exhaustive comparison among
matrix-completion algorithms. In the extended version, we will evaluate additional
matrix-completion methods under the same masking protocol and compare their
reconstruction accuracy, robustness, and convergence behavior.

Overall, this study provides practical guidance for selecting imputation
strategies that preserve the integrity of performance portability analyses.
By making the treatment of missing data explicit and principled, researchers
can avoid misleading conclusions and enable more reliable comparisons across
platforms, programming models, and architectures.

\bibliographystyle{splncs04}
\bibliography{PPAM_2026_refs} 

\end{document}